\documentstyle[multicol,epsf,aps]{revtex}
\begin{document}
\def\T{\Theta} 
\def\D{\Delta}
\def\d{\delta}
\def\r{\rho}
\def\p{\pi}
\def\a{\alpha}
\def\g{\gamma}
\def\ra{\rightarrow}
\def\s{\sigma}
\def\b{\beta}
\def\e{\epsilon}
\def\G{\Gamma}
\def\o{\omega}
\def\pe{$1/r^\a$ }
\def\l{\lambda}
\def\f{\phi}

\title{Collapse in $1/r^\a$ interacting systems }
\author{I.~Ispolatov
and E.~G.~D.~Cohen}
\address{Center for Studies in Physics and Biology, Rockefeller University, 
1230 York Ave, New York,
NY 10021, USA.} 
\maketitle

\begin{abstract} 
\noindent  
Collapse, or a gravitational-like phase transition is studied in a 
microcanonical ensemble
of particles with an attractive $1/r^\a$ potential.
A mean field continuous integral equation is used to determine a saddle-point 
density
profile that extremizes the entropy functional. 
For all $0<\a<3$, a critical energy is determined below which
the entropy of the system exhibits a discontinuous
jump. If an effective short-range cutoff is applied, the entropy jump
is finite; if not, the entropy diverges to $+\infty$.
A stable integral equation solution represents a state with 
maximal entropy; the reverse
is always true only for a modified integral equation introduced here.
\medskip\noindent{PACS numbers: 02.30.Rz 04.40.-b 05.70.Fh 64.60.-i}

\end{abstract}

\section{Introduction}
The behavior of systems with long-range interactions is often
different from those considered in traditional thermodynamics.
As an example we take self gravitating systems, i.e. particle ensembles
with purely attractive $1/r$ interactions. It is known
that such systems exhibit collapse, sometimes called a zero-order
phase transition, when the energy in the microcanonical ensemble (ME) or the 
temperature
in the canonical ensemble (CE) drop below a certain critical value $e_c$
or $T_c$, respectively
\cite{an,dv,pr}. During such a transition, the corresponding thermodynamic 
potentials
(entropy in the ME or free energy in the CE) exhibit a discontinuous jump.
If the interaction between the particles is purely attractive and no 
short-range
cutoff is introduced, then the discontinuous jump is infinite and the 
entropy and free energy go to $+\infty$ and
$-\infty$, respectively. This makes all normal (noncollapsed) states of 
the self-attractive system metastable
with respect to such a collapse; the collapse energy $e_c$ is in fact an 
energy 
below which the
metastable state ceases to exist. 
If, on the other hand, some form of  short-range cutoff is introduced, 
the entropy and
free energy jumps are finite. In this case as a result of the collapse, 
the system
goes into a nonsingular state with a dense core, the precise nature of
which depends on the
details of the short-range behavior of the potential.
Then only the normal states which are in some interval of 
energies above the collapse point 
are metastable with respect to such a transition (see Fig.~1). 
\begin{figure}
\centerline{\epsfxsize=8cm \epsfbox{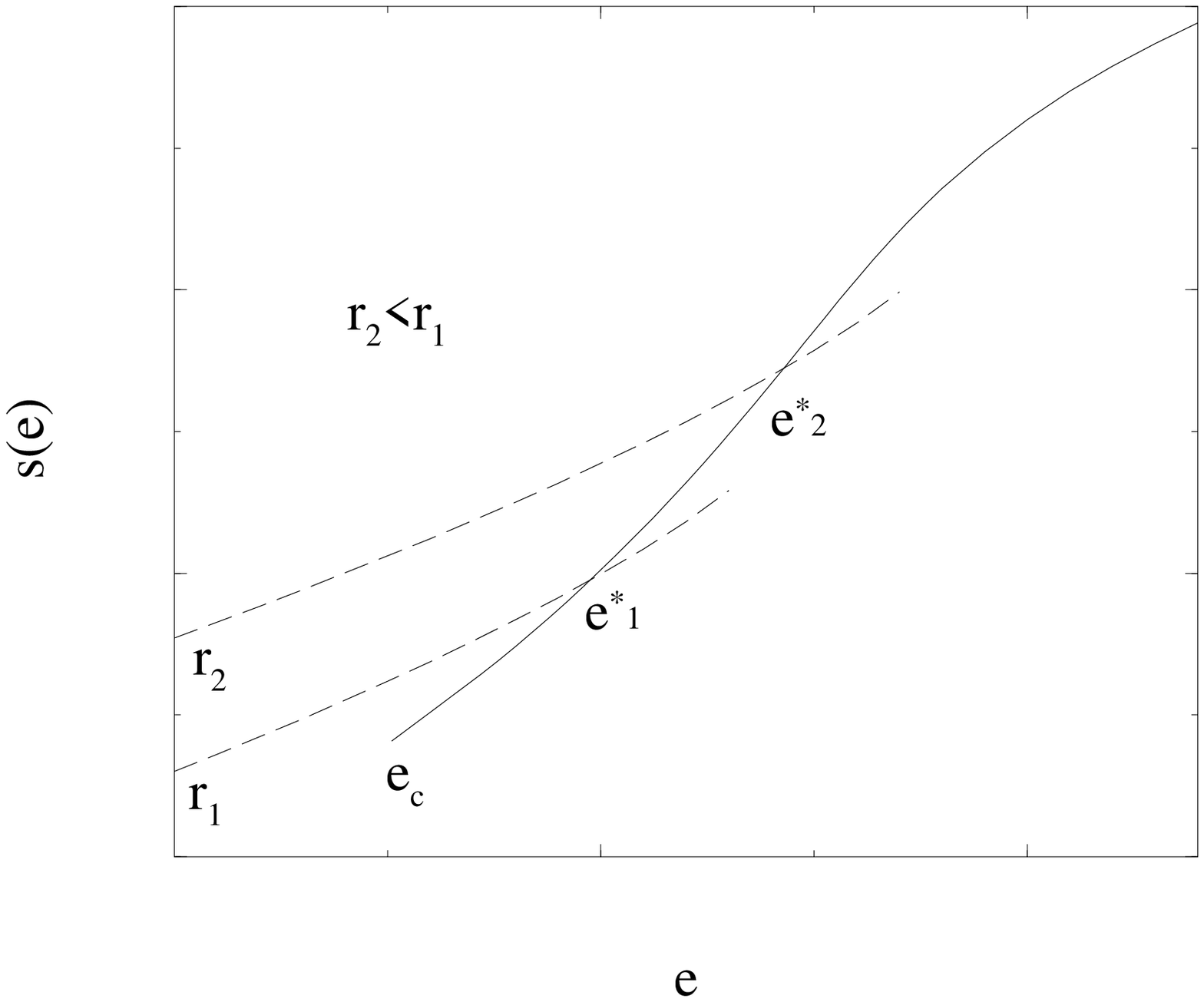}}
\noindent
{\small {\bf Fig.~1.}
Sketch of an entropy vs. energy plot for a system with
gravitational-like collapse. The entropy  of the normal (non-collapsed)
state is shown by a solid line, the entropies of the two collapsed states
for different cutoff radio $r_1$ and $r_2$, $r_2<r_1$, are shown
by dashed lines. The entropies of the two collapsed
states intersect the entropy  of the normal state at energies
$e_1^*$ and $e_2^*$.
}
\end{figure}
There is
an energy $e^*$ for which both collapsed 
and normal system have the same entropy (cf. $e_1^*$ and $e_2^*$ in Fig.~1);
above this energy the collapsed state becomes 
metastable and at some higher energy it ceases to exist \cite{bm,bm2}.
It is possible to regard the energy $e^*$ 
as that where a true phase transition occurs. 
When the effective cutoff vanishes, 
such phase transition energy $e^*$ increases to infinity, so that
without a cutoff all the finite energy states are metastable \cite{ki}.
However, the value of $e^*$ is highly sensitive
to the details of the short-range cutoff. On the contrary, the collapse energy 
$e_c$ depends on the long-range part of the interparticle interactions and
is almost unaffected by a cutoff, provided that it is sufficiently short-range.
Since the collapse phenomena carry most of the information
about the
peculiarities of the phase behavior of the system caused by
the long-range interactions, they usually gain the most attention, considering
then a phase transition at $e^*$ as a mere artifact of an often auxiliary 
cutoff.

While these rather elaborate studies of gravitational collapse have usually 
been 
motivated by
cosmological applications and were performed solely for an $1/r$ potential,
a natural question is, what happens if the particles interact via an 
attractive \pe potential with arbitrary $\a$.
It has been noticed before that in systems with nonintegrable
interactions, i.e. when $\a$ is less than the dimensionality of the space,
first order phase transitions occur differently in the ME than in the CE,
even for
$N\ra\infty$ \cite{us}. However, in the examples considered in the literature,
the
potential energy was always bounded from below (usually by putting the system
on a lattice), allowing only for normal first-order phase transitions and
excluding any singular collapse. 

In this work we consider a possibility of collapse in systems, 
similar to the gravitational
self-attracting Hamiltonian particle systems, 
but with a general \pe potential. 
The paper is organized as follows:
after this brief introduction we formally define the 
model and derive an integral equation to solve for a saddle-point density
profile.
Then we describe the results obtained by a numerical iterative solution 
of this 
equation for various $0<\a<3$. After this, we look at the local stability of 
the solutions of the integral equation and link their stability
to the stability (metastability) of the corresponding states of the 
self-attracting system. Finally, we discuss the results and
directions for further investigation.

While the collapse is shown to be a generic property
of self-gravitating systems occurring in both the ME and the CE, the 
ME allows to 
obtain more information
about the system than the CE. For example, negative heat-capacity states, that 
precede the collapse in the ME, are not accessible in the CE, as it is in the
much better studied case of normal first-order phase transitions 
\cite{hu1,us,gr}.
In the rest of our paper we consider systems only in the
ME.
In our studies we strongly rely on the large body of existing results derived
for $\a=1$, often rigorously \cite{dv,pr,ba}.

\section{The Model}
Let us consider a 3-dimensional ME of $N$ particles 
each having mass $m$, 
interacting with an attractive pair potential $V_{ij}= -G/|\vec r_i-
\vec r_j|^{\a}$. These particles are confined to a spherical
container with radius $R$; the total energy of the particles is $E$.
The microcanonical entropy (equal to the logarithm of the density of states; 
here and in the following we put $k_B=1$)
can be expressed as:
\begin{equation}\label{s1}
S(E)  = \log\left\{ {1 \over N !}\int\ldots \int
\prod_{k=1}^N{{d \vec p_k\, d \vec r_k}\over{(2\pi\hbar)^3}}\; 
\delta\!\left[E - \sum_{l=1}^N\;{{p_l^2}\over{2m}} - 
\sum_{i=1}^N \sum_{j=1+1}^N V_{ij}\right] \right\} 
\end{equation}

After integrating over the momenta and introducing a dimensionless energy
$\e= E R^{\a}/GN^2$ and radial coordinates $x_i\equiv r_i/R$, 
Eq.~(\ref{s1}) can be reduced to:

\begin{equation}
\label{s2}
S(\e)=S_0+\log\left[
\int\ldots \int
\prod_{k=1}^N d \vec x_k\ \; 
\Theta\:(\e_{kin})\:  
\e_{kin}^{3N/2-1}\right], 
\end{equation}
where $\e_{kin}$ is a dimensionless kinetic energy,
\begin{equation}
\label{ek}
\e_{kin}=\e + {1\over N^2}
\sum_{1\leq i<j\leq N} {1\over |x_i-x_j|^{\a}}
\end{equation}
and $S_0$ is an energy-independent term,
\begin{equation}
\label{s3}
S_0  = \log\left[ 2m \left({m \over 2\p\hbar^2} \right)^{3N/2} \!
\left({G N^2 \over R^{\a}}\right)^{3N/2-1} \!
{R^{3N}\over \G({3N/2}) N!}\right].
\end{equation}
Here $\G(x)$ is a Gamma function, $\T(x)$ is a unit step function which
guarantees positiveness of the kinetic energy,
and the integration for each $d\vec x_i$ runs over a 3-dimensional unit sphere.
It follows from Eq.~(\ref{s2}) that the entire thermodynamical behavior
of our system depends on the single variable $\e$; 
and for a system to remain in the same state when the number of particles is 
varied but the energy per particle is fixed, the system size $R$ should scale
as $R \sim N^{1/\a}$. This indicates that $\a=3$ is the 
largest value of $\a$ for which no thermodymanic scaling
is observed and hence some non-traditional phase transitions may exist.

We assume that the number of particles $N$ is large and go to a continuum 
limit,
replacing a $3N$-dimensional configurational 
integral in (\ref{s2})
by a functional integral over possible density profiles $\r(\vec x)$
(see, for example, \cite{dv,pr}):
\begin{equation}
\label{s4}
S(\e)\approx S_0+\log\left\{
\int {\cal D}\r \;\int_{-i\infty}^{+i\infty} {d\g \over 2 \pi i}\;
\int_{-i\infty}^{+i\infty} {d\b \over 2 \p i}\;
\exp(N \tilde s[\r(.),\e,\g,\b])\right\},
\end{equation}
where the ``effective action'' functional $\tilde s$ is defined:
\begin{eqnarray}
\label{s5}
\nonumber
\tilde s[\r(.),\e,\g,\b]=
\b\left(\e+{1\over 2}\int\int{\r(\vec x_1)\r(\vec x_2)\over
|\vec x_1-\vec x_2|^{\a}}d\vec x_1 d\vec x_2 \right)+\\
\g \left(\int\r(\vec x)d\vec x -1 \right)-{3\over 2}\ln \b-
\int\r(\vec x)\ln{\r(\vec x)\over e} d\vec x.  
\end{eqnarray}
Here we introduced two auxiliary Fourier integrations: one
over $d\b$ to replace the $\T$ 
function in Eq.~(\ref{s3}) (see e.g. \cite{dv}),
\begin{equation}
\nonumber
x^{\s}\T(x)={\G(\s+1)\over 2\pi}\int_{-\infty}^{+\infty}e^{i\o x}
{d\o\over (i\o)^{\s+1}},
\end{equation}
and similarly one over
$d\g$ to express the $\d$ function 
$\d \left[ \int\r(\vec x)d\vec x -1 \right]$. 
The Eq.~(\ref{s4}) is applicable only for $0<\a<3$ when the 
integrals on $d\vec x_1$ and $d\vec x_2$ are convergent at the lower limit.

Using that $N$ is large, the integral in Eq.~(\ref{s4}) can be evaluated by 
the saddle-point method. To determine the dominant contributions
to the integral, we differentiate (\ref{s5}) with respect to
$\g,\; \b$, and $\r(\vec x)$ 
and look for $\{\g_s, \b_s , \r_s(\vec x)\} $ which give  the extrema to the 
effective action (\ref{s5}).
As a result, unique solutions are obtained for $ \{\g_s,\b_s \} $; while
for $\r_s(\vec x) $ a nonlinear integral equation is obtained
for which the number and the nature of solutions is generally unknown:

\begin{eqnarray}
\label{extr}
\nonumber
\g_s=-\ln\left\{\int \exp\left[ {\b_s\r_s(\vec x_1) 
\over|\vec x_1-\vec x_2|^{\a}} d \vec x_1 \right ] d \vec x_2 \right \}\\
\b_s={3\over 2} \left[\e + {1\over 2}\int \int{\r_s(\vec x_1) \r_s(\vec x_2) 
 \over |\vec x_1-\vec x_2|^{\a}}d \vec x_1 d \vec x_2\right]^{-1}\\
\nonumber
\r_s(\vec x)=\exp \left [\g_s+\b_s \int {\r_s(\vec x_1) 
\over |\vec x_1-\vec x|^{\a}} d \vec x_1 \right ]
\end{eqnarray}
It follows from (\ref{extr}) that $\b_s$ is equal to 3/2
of the kinetic energy, hence it is equal to the
entropy derivative with respect to the energy, i.e. to the 
inverse temperature, $\b_s=ds(\e)/d\e\equiv1/T$.

Using (\ref{s5},\ref{extr}) and ignoring terms independent of  $\e$, 
we obtain for the 
entropy per particle $s(\e)=S(\e)/N$ in the saddle-point approximation:
\begin{equation}
\label{s0}
s(\e)={3\over 2} \ln \left[\e+{1\over 2}\int \int{\r(\vec x_1) \r(\vec x_2) 
 \over |\vec x_1-\vec x_2|^{\a}}d \vec x_1 d \vec x_2 \right ]
-\int\r(\vec x)\ln[\r(\vec x)] d \vec x + {\cal O}({1\over N}).
\end{equation}
The saddle-point approximation works only if the second derivatives
of the effective action functional $\tilde s$ are not too small; 
we will return to the question of its validity in Section IV.

The above equations look similar to those for the gravitational ($\a=1$) 
case which are derived, for example, in \cite{dv,pr}, to where the reader is 
referred for a more detailed description of the derivation.
However, we can not proceed further along the lines developed in the 
references mentioned above, which all deal with the gravitational case.  
For, unlike the gravitational case,
when the $\Delta (1/r)=-4\pi\d(r)$ property of a Coulomb 
interaction $1/r$
allows one to reduce
(\ref{extr}) to a second-order differential equation, 
no such property exists for a general $\a$, and we have to deal with the
integral equation for $\r_s(\vec x)$ in (\ref{extr}) directly.

In the following a further simplification is made by disregarding  
the angular dependence for the saddle-point density profile, i.e.
$\r(\vec x)=\r(x)$; for the gravitational case this is justified
in \cite{dv,pr}. After performing a straightforward
angular integration in the exponential of Eq.~(\ref{extr}), 
we obtain for $\a\neq2$,
\begin{eqnarray}
\label{main}
\nonumber
\r(x)=\r_0\exp \left\{ {2\pi \b_s\over (2-\a) x} \int_0^1 \r(x_1) x_1 
\left[ (x+x_1)^{(2-\a)}-|x-x_1|^{(2-\a)} \right] d x_1 \right\},\\
\r_0=\left\{ 
\int_0^1 \exp \left[ {2\pi \b_s\over (2-\a) x'} \int_0^1 \r(x_1) x_1 
\left( (x'+x_1)^{(2-\a)}-|x'-x_1|^{(2-\a)} \right) d x_1 \right] dx'
\right\}^{-1},\\
\nonumber
\b_s={3\over 2} \left\{\e + {4\pi^2\over (2-\a) }
\int_0^1 \int_0^1 \r(x_1) \r(x_2) x_1 x_2 \left[ (x+x_1)^{(2-\a)}-
|x-x_1|^{(2-\a)} 
\right] d x_1 d x_2 \right\}^{-1};
\end{eqnarray}
and for $\a=2$,
\begin{eqnarray}
\label{main2}
\nonumber
\r(x)=\r_0\exp \left\{ {2\pi \b_s\over  x} \int_0^1 \r(x_1) x_1 
\left[\ln (x+x_1) - \ln|x-x_1| \right] d x_1 \right\},\\
\r_0=\left\{ \int_0^1 \exp \left[ {2\pi \b_s\over  x'} \int_0^1 
\r(x_1) x_1 
\left( \ln (x'+x_1)-\ln |x'-x_1| \right) d x_1 \right] dx' \right\}
^{-1},\\
\nonumber
\b_s={3\over 2} \left\{\e + 4\pi^2
\int_0^1 \int_0^1 \r(x_1) \r(x_2) x_1 x_2 \left[\ln (x+x_1) -
\ln |x-x_1| 
\right] d x_1 d x_2 \right\}^{-1}.
\end{eqnarray}.

Once the Eqs.~(\ref{main}, \ref{main2}) have been solved and the
entropy (\ref{s0}) has been calculated, the nature of the phase 
behavior of the system can be deduced from an entropy-energy plot.
In the next section we analyze the Eqs.~(\ref{main}, \ref{main2}) numerically.

\section{numerical solution}

If $\b_s$ is considered an independent parameter rather than
a factor depending on $\e$ and $\r_s$, the Eq.~(\ref{extr}) is often called
a Generalized Poisson-Boltzmann-Emden Equation \cite{ba}.
Very little is known about this equation
even in the gravitational $\a=1$ case. The only exactly known, 
so called ``singular'', solution is for $\b_s=2$ \cite{ba} and 
has the form $\r_{sing}(x)=(4\pi x^2)^{-1}$, which leads to $\e=-1/4$ and
$s=\ln(\sqrt{2}\pi) -2$. 
However, we are interested here in general $\a$
and $\e$; and $\b_s$ in the ME
is not an independent variable but a function
of $\e_s$. As we will discuss below, this  has a crucial effect on
the numerical accessibility of solutions in a certain interval of $\e$.

To solve the Eqs.~(\ref{main}, \ref{main2}) numerically 
for general $\a$ and $\e$, we use
a simple iterative method. For a fixed  $\a$ we start at a relatively high
value of $\e$ with a flat density profile, $\r_0=3/4\pi$, which is the 
solution of (\ref{main}, \ref{main2}) for $\e\ra+\infty$.
Putting $\r_0$ into (\ref{main}, \ref{main2}), we calculate 
$\b_s$ and obtain a new density profile $\r_1(x)$. 
In other words, an iterative map
\begin{equation}
\label{it}
\r_{i+1}(x)=F_{\e}[\r_i(.),x]
\end{equation}
is introduced, with a nonlinear functional $F_{\e}[\r_i(.),x]$ defined
by (\ref{main}, \ref{main2}). 

After a sufficient convergence of the iterations (\ref{it}) is achieved,
i.e., when
\begin{equation}
\label{con}
4\pi\int_0^1|\r_{i+1}(x)-\r_{i}(x)|x^2dx < \d<<1,
\end{equation}
the entropy is calculated with (\ref{s0}). We then move to a lower energy 
point, 
use the previous energy point
density profile as $\r_0$, and repeat the procedure again.
Caution in selecting the initial energy as well as the energy step
has to be exercised to maintain the positiveness of the kinetic energy.
In fact the 
initial energy has to be larger than $-{3^2 2^{2-\a}\over (6-\a)(4-\a)(3-\a)}$,
which is the potential energy for the flat density profile.
As we progress towards lower energies, the density profiles get more and more 
peaked near the origin, and the absolute value of the potential energy
increases.
To allow for improper integrals in the  numerical integration of 
(\ref{main}, \ref{main2}), 
we use a simple mid-point trapezoid rule. 
Uniform meshes
of 1000-2000 points were usually sufficient. However, in order 
to determine the position of
a phase transition point with  a sufficiently high precision in order to
compare our results to the existing ones for $\a=1$, and to
achieve also a discontinuous phase transition for smaller $\a$, we had to use
finer meshes with up to 5000 points. The convergence parameter
$\d$ was usually set to be $10^{-6}$.

The main conclusion that can be derived from the numerical results obtained is 
the following:
for all $0<\a<3$, as for $\a=1$, there is a certain energy $\e_c(\a)$ 
below which the system collapses and the entropy exhibits a discontinuous jump.
The results for $\e_c(\a)$ are presented in Fig.~2.
\begin{figure}
\centerline{\epsfxsize=8cm \epsfbox{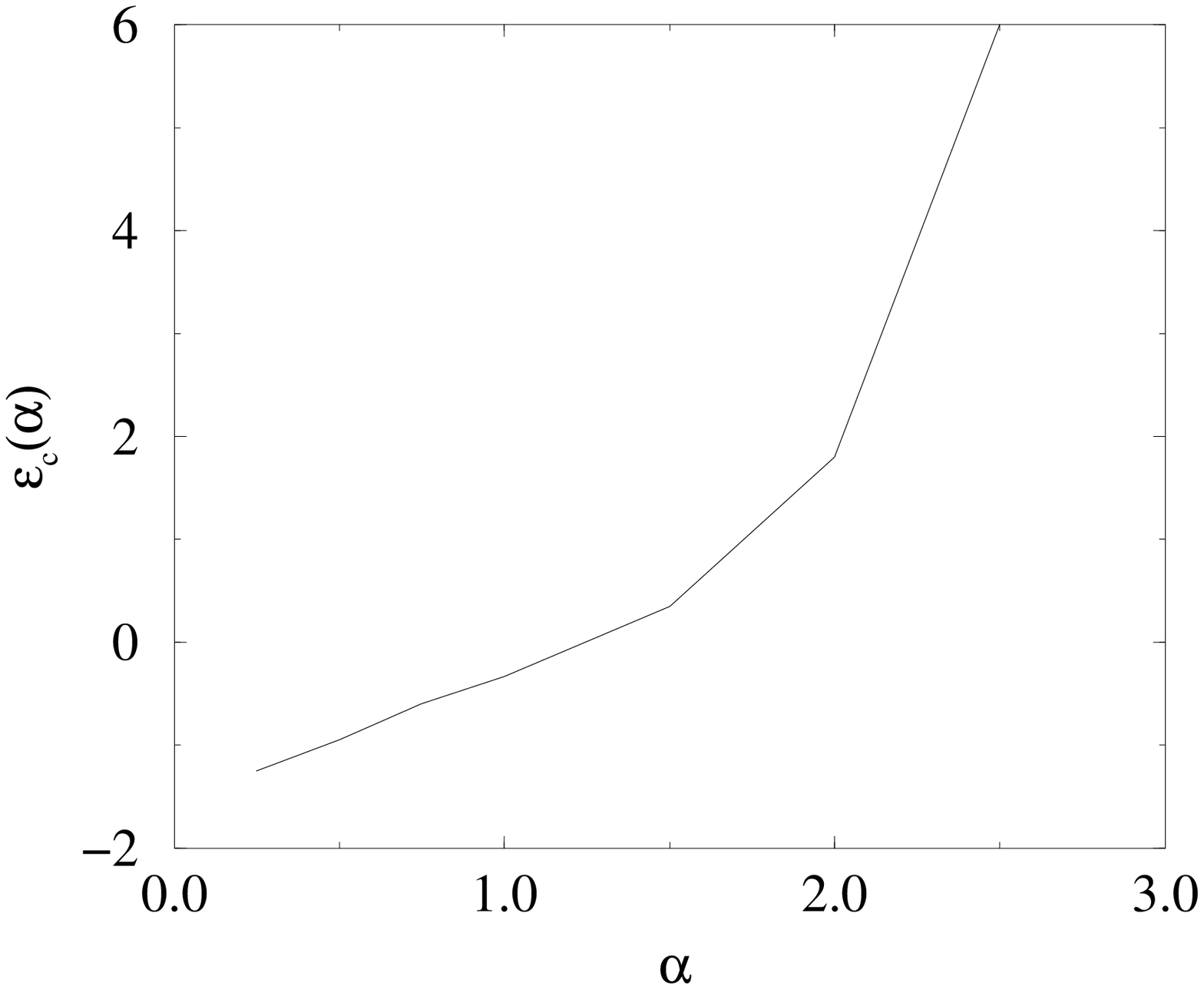}}
\noindent
{\small {\bf Fig.~2.}
Plot of collapse energy $\e_c(\a)$ vs. potential exponent $\a$.
}
\end{figure}
To verify our calculations of $\e_c(\a)$, we compare our result for 
$\e_c(\a=1)$ with the 
existing data obtained by other methods. Our number, $\e_c(\a=1)=-0.3346$, 
is consistent with $\e_c(\a=1)=-0.335$, quoted in \cite{pr,bm}.

Since the behavior of self-attractive systems is quantitatively similar 
for all $0<\a<3$,
let us consider in more detail, for example,
a system with $\a=1/2$. Plots of entropy $s(\e)$ and
inverse temperature $\b_s(e)=ds(\e)/d\e$ are presented in
Fig.~3. 
\begin{figure}
\centerline{\epsfxsize=8cm \epsfbox{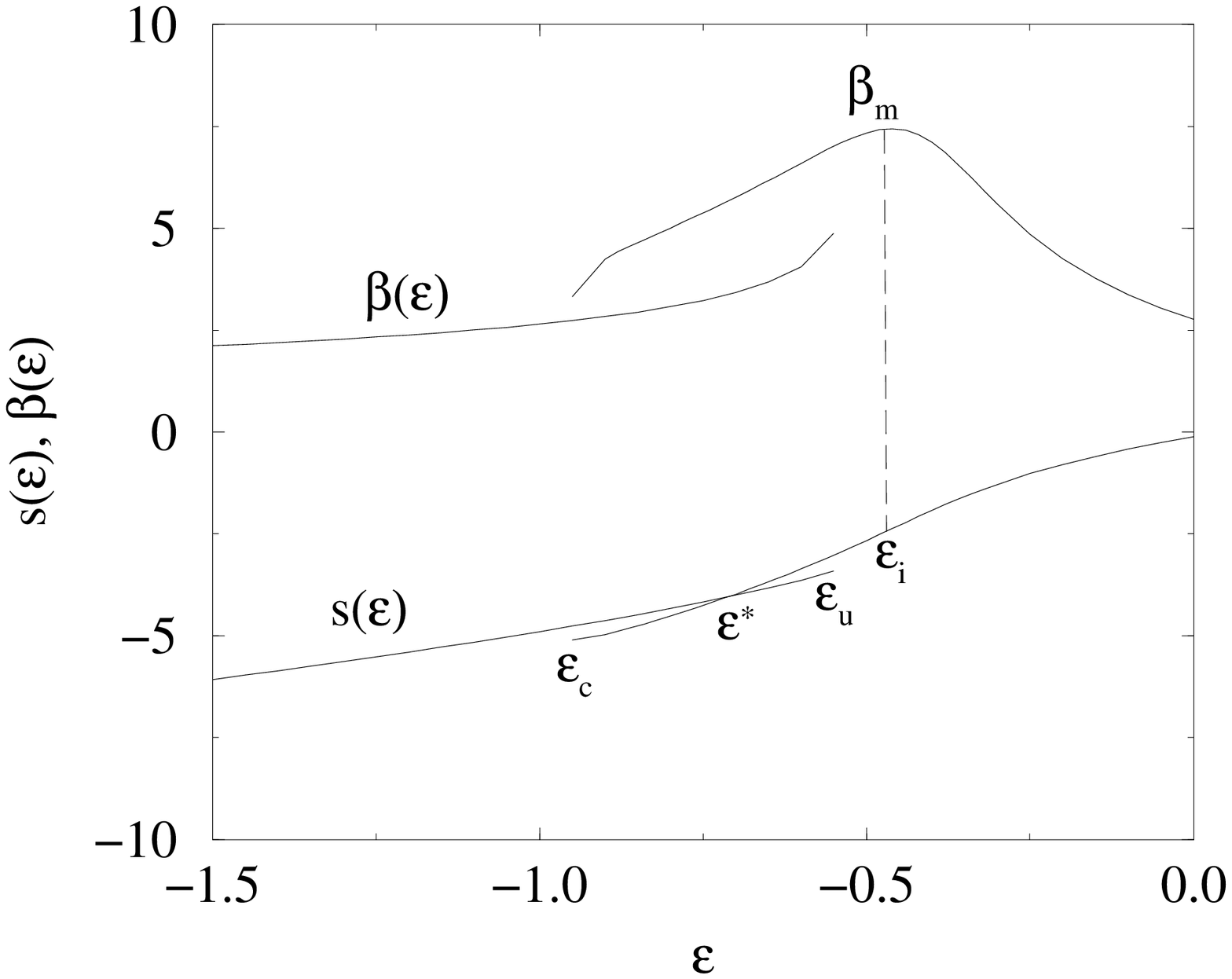}}
\noindent
{\small {\bf Fig.~3.}
Plots of entropy $s(\e)$ and entropy derivative $\b_s(\e))$
for $\a=1/2$. The radius of excluded central volume
$r_0=5\times 10^{-4}$. The points $\e_c, \e^*, \e_u$, and $\e_i$ are defined
in the text.
}
\end{figure}

As we go down along the energy axis $\e$, the entropy decreases
as well, passing through an inflection point $\e_i$ where $\b$ reaches
its maximum $\b_m$. For energies below this inflection point, the
system has a negative specific heat ($d^2s(\e)/d\e^2>0$) and is therefore
unstable in the CE.
As we pass through the $\e_i$ point and continue decreasing the energy, the 
convergence of (\ref{it}) becomes slower and slower, and at the
point $\e_c$ the iterations start to diverge. It is straightforward to
show for all $0<\a<3$ (see, e.g. \cite{ba} for $\a=1$) 
that the entropy is unbounded from above with respect
to uniform squeezing of all the matter into a sphere with a 
radius going to zero.
Hence, if no short-range cutoff is present, it is reasonable to assume
that the entropy discontinuity at $\e_c(\a)$ is infinite.

If some form of a short-range cutoff is introduced, the
entropy discontinuity may become finite. To investigate this
we tried two approaches. One, suggested in \cite{bm}, is to place a small
spherical excluded volume with a radius $r_0$ in the center of the system, 
or, in other words,
to replace a spherical container with a spherical shell container.
The other approach is to replace the original ``bare'' potential
$1/r^{\a}$ with a 
``soft'' potentials of the form $1/(r^2+r_0^2)^{2\a}$.
For a reasonably small short-range cutoff ($r_0\sim 10^{-3}$ for small $\a$, 
$r_0\sim 10^{-2}$ for $\a\simeq 3$ for both approaches) 
the behavior of the pre-collapsed system is virtually unaffected.
A finiteness of the integration mesh can also play the role of
an effective central excluded volume; 
for our method of integration the size of such an effective 
excluded volume is roughly of the order of the mesh step. 

Introduction of a short-range cutoff makes the existence of a
non-singular collapsed phase possible. However, being applied directly,
the iterative method (\ref{it}) still diverges when $\e<\e_c$. To make
it convergent, we introduced a map  with a variable ``step'',
\begin{equation}
\label{it1}
\r_{i+1}(x)=\s F[\r_i(.),\e]+(1-\s)\r_i(x),
\end{equation}  
where $0<\s\leq 1$ is the step size parameter. Choosing $\s$ sufficiently small
(as small as  $\sim 10^{-2} - 10^{-3}$), 
we were able to make the algorithm convergent
for $\e<\e_c$. The connection between the numerical 
stability of the 
iterative algorithm and the thermodynamic stability of the 
corresponding phase is analyzed in the next 
section. A typical density profile in the collapsed phase  
exhibits a much higher concentration around the origin than
to the normal (un-collapsed) phase; plots of density profiles for $\a=1/2$
are presented in Fig.~4.
\begin{figure}
\centerline{\epsfxsize=8cm \epsfbox{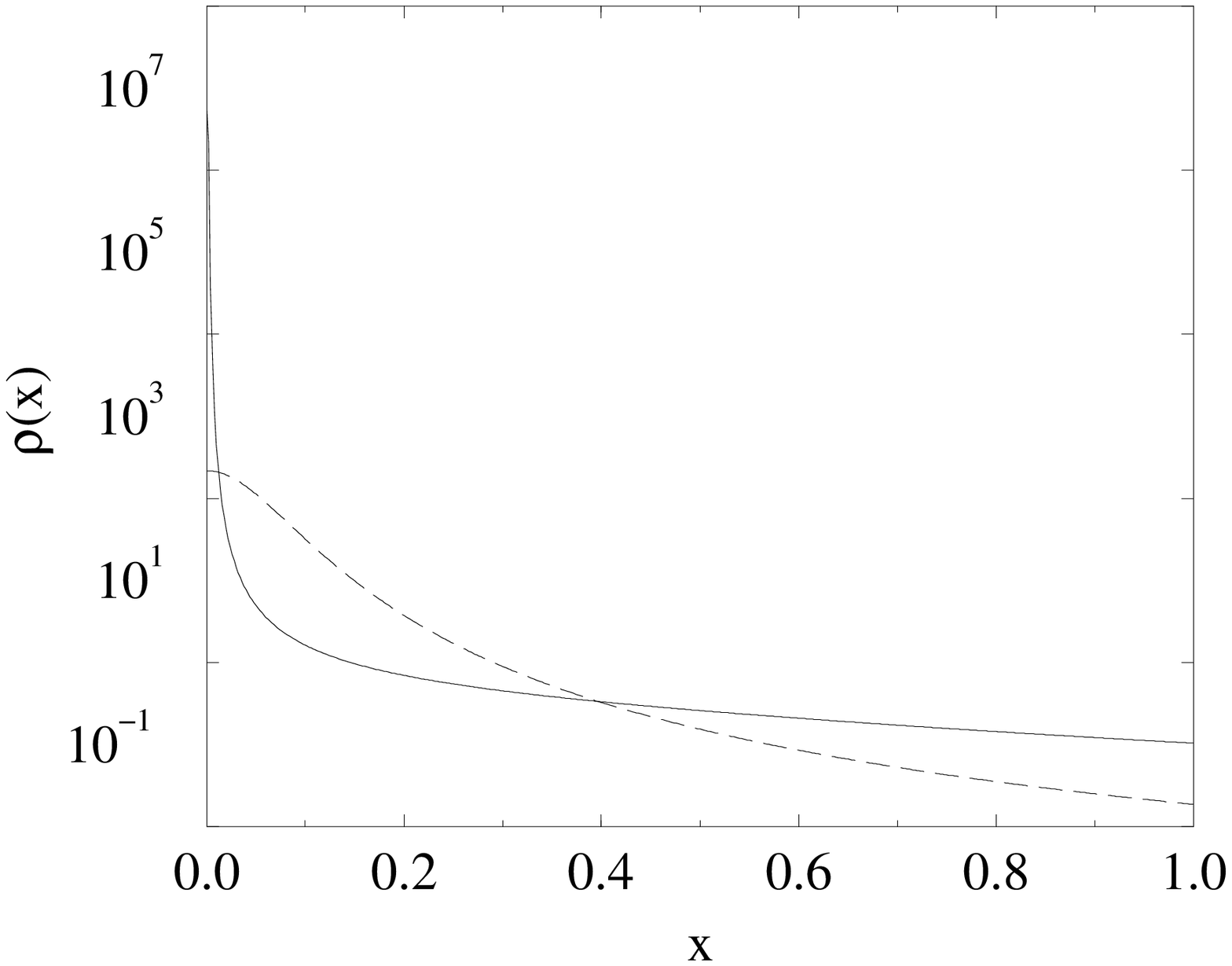}}
\noindent
{\small {\bf Fig.~4.}
Density profiles $\r(x)$ for $\a=1/2$ for
normal (dashed line) and collapsed (solid line) phases
for the energy $\e^*=-0.708$, when entropies of both
phases are the same. The radius of the excluded central volume is
$r_0=5\times 10^{-4}$.
}
\end{figure}
A collapsed phase exists not only for $\e<\e_c$, but for
$\e>\e_c$ as well. In fact, this phase is globally stable in the range of 
energies where its entropy is
higher than that of the normal phase, i.e. when $\e<\e^*$.
For $\e>\e^*$, the collapsed phase is metastable and above
some energy $\e_u$ becomes unstable even locally (see Fig.~3). However, 
being cutoff-dependent,
$\e^*$ and $\e_u$ are not directly related to the fundamental properties of
the original \pe self-attracting system.

Finally we return to the exact $\r_{sing}(x)=(4\pi x^2)^{-1}$ solution
which exists for $\e=-1/4$ and $\a=1$. Our attempts to
approach this solution by the numerical iterative methods (\ref{it},\ref{it1}) 
failed. In fact, even after
substituting the $\r_{sing}(x)$ into (\ref{it1}) as an initial approximation 
$\r_0(x)$, the iterative solution of (\ref{it1}) evolved either to a 
normal or to a 
collapsed solution depending on the value of the step $\s$.
We calculated the entropies for
the three solutions, that exist at $\e=-1/4$: a normal $s_n$, a collapsed $s_c$
(with a sufficiently small central core, 
so that it still exists at this energy), and a $s_{sing}$ for $\r_{sing}$. 
It turns out that $s_{sing}<min\{s_n,s_c\}$, which, together with
the evidence obtained from the iterative procedures mentioned above, 
strongly suggests
that in the space of solutions (or fixed points) of (\ref{it1}),
both  normal and collapsed $\r(x)$ are at least locally stable
(attractive), while $r_{sin}$ is unstable (repulsive).

\section {stability of the thermodynamic state and the iterative map} 
The iterative solutions of the integral equation (\ref{main},\ref{it})
for the saddle point density profile $\r_s(x)$ of the self-attracting system
correspond to thermodynamically stable or unstable states.
In this section we will give the necessary and the sufficient conditions
for the thermodynamic stability of the  $\r_s(x)$ in terms of the stability
of the iterative solutions of the integral equation (\ref{main}).
To that end we will look at the stability of  (\ref{it}) for a certain
trial function $\r_i(x)$.
Let us assume that $\r_s(x)$ is a solution (fixed point) of 
(\ref{it}), and $\d\r_i(x)=\r_i(x)-\r_s(x)$ is a small deviation
of the {\it i}th iteration from the solution $\r_s(x)$. 
After one iteration we obtain
for $\d \r_{i+1}(x)=\r_{i+1}(x)-\r_s(x)$
\begin{equation}
\label{1v}
\d \r_{i+1}(x)=\int_0^1 
\left.{\d F_{\e}[\r(.),x]\over \d \r(x')}\right|_{\r=\r_s}
\d\r_i(x')dx' +{\cal O}(\d\r^2)
\end{equation}
For the iterations to converge, $|\d \r_{i+1}(x)|<|\d \r_{i}(x)|$ for all $x$.
This condition is equivalent to a requirement that the absolute values of
all the eigenvalues $\l_i$ of the first variation operator defined below 
are less than one,
\begin{equation}
\label{eig}
\l_i\:\f_i(x)=\int_0^1
\left.{\d F_{\e}[\r(.),x]\over \d \r(x')}\right|_{\r=\r_s}
\f_i(x')dx',
\end{equation}
where the $\f_i$ are the eigenfunctions corresponding to $\l_i$.
Recalling the definition of $ F_{\e}[\r(.),x]$ 
(\ref{extr},\ref{main},\ref{main2}) we note that it can be expressed
through a first variation of the effective action functional
$\tilde s[\r(.),\e,\g,\b]$:
\begin{equation}
\label{sf}
 F_{\e}[\r(.),x]=\exp\left[\left.{\d\tilde s[\r(.),\e,\g,\b]\over d\r(x)}
\right|_{\g=\g_s,\b=\b_s}+\ln\r(x)\right].
\end{equation}
Then one obtains for the first variation of $F_{\e}[\r(.),x]$:
\begin{equation}
\label{sfm}
\left.{\d F_{\e}[\r(.),x]\over \d \r(x')}\right|_{\r=\r_s}=
\r_s(x)\left.{\d^2\tilde s[\r(.),\e,\g,\b]\over \d\r(x)\d\r(x')}
\right|_{\g=\g_s,\b=\b_s,\r=\r_s}+\d(x-x').
\end{equation}  
Now for a state $\r_s(x)$ to be thermodynamically stable and
for the saddle point approximation to be applicable at all, the second
variation of the effective action $\tilde s$ on the right hand side of
(\ref{sfm}) 
must be negative for all $x$ and $x'$.
This is equivalent to the requirement
that all the eigenvalues of the operator on the right-hand side of
(\ref{sfm}) are less than one. For,
the only eigenvalue of the delta-function $\d(x-x')$ is one with any
function being its eigenfunction and the density $\r_s(x)$ is strictly positive
for all $0\leq x\leq 1$. Hence, the convergence of the
map (\ref{it}) to the function $\r_s(x)$ is a sufficient condition 
for the thermodynamic stability 
of $\r_s(x)$ and the validity of the saddle point approximation. 
However, the reverse is not true, i.e. stable thermodynamic states do not 
necessarily correspond to stable iterative solutions. For,
the operator on the left-hand
side of Eq.~(\ref{sfm}) may have eigenvalues which are less than -1, which will
make the map (\ref{it}) unstable. Therefore let us instead of (\ref{it}) 
consider the
variable step map (\ref{it1}) with eigenvalues of its first variation 
$\l_i^*(\s)=\s\l_i+(1-\s)$. Then for any negative $\l_i$ we can chose
an appropriate $\s$ so that $|\l_i^*(\s)|<1$. Evidently, this is what 
happened in the case of
the collapsed phase which, possessing the highest entropy, is definitely 
stable,
but can be accessed numerically only by using  (\ref{it1}) with a sufficiently 
small $\s$. On the other hand, for thermodynamically unstable states, 
such as $\r_{sing}$,
some of the eigenvalues of the operator in the right-hand side of
Eq.~(\ref{sfm}) are larger than 1, which make the iterative
maps (\ref{it},\ref{it1}) unstable, so that such solutions cannot be
found iteratively. This completes the demonstration of the equivalence of the 
two stabilities.

\section{discussion}
1. In the previous sections we revealed the existence of collapse and
associated with it a 
discontinuity in the entropy 
in the microcanonical ensemble
of particles with $1/r^{\a}$ attraction for $0<\a<3$. 
This discontinuity was an infinite jump if no short-range cutoff was 
present.
A carefully introduced short-range cutoff leaves the properties of
pre-collapsed system virtually unaffected, but makes the entropy jump finite
and allows to observe the collapsed phase. We proved that the stability
of a solution of the integral equation for a saddle-point density profile  
is a sufficient condition for the profile to make the entropy a maximum and 
therefore be either stable or metastable. We modified the integral equation to 
make the reverse also true.

2. In the range of $\a$ we have been working with, $0<\a<3$,
the potential is often called ``non-integrable'', since the integral
$\int d^3r/r^{\a}$ diverges at its upper limit. 
As the potential becomes integrable ($\a>3$),  
the continuum approach used here becomes
inapplicable because the short-range density fluctuations,
which the continuous approach cannot account for, become
dominant over the long-range effects.
Formally, the short-range nature of the behavior of the systems for
$\a>3$ manifests itself as the divergence of the
integral $\int d^3r/r^{\a}$ at its lower limit.

3. A very important question which remains is that of
the order of the gravitation-like phase transition.
Here we have to distinguish between the collapse itself, which
happens at $\e_c$, and the phase transition which happens
at the energy $\e^*$ where the entropies of non-collapsed
and collapsed states are equal (see Fig.~3).
Since the entropy at the collapse point $\e_c$ exhibits a discontinuous
jump, the collapse is often called a zero-order phase transition \cite{dv}.
However, the collapse is not a phase transition in the formal sense
since it converts a metastable state into a stable one, which
can be singular or finite, depending on the presence of the short-range cutoff.

On the other hand, the ``true'' phase transition between stable phases, 
which happens at $e^*$, is sometimes referred to as
a ``gravitational first-order phase transition'' \cite{bm2}. Its
distinct features include an inability of the two phases 
(non-collapsed and collapsed) to coexist and a discontinuous
$\b(\e)$ i.e. temperature  \cite{bm2}.
Yet in the ``normal'' ME first-order phase
transition in a long-range interacting system 
(such as a mean-field Potts model),
$\b((\e)$ remains continuous and smooth,
but exhibits non-monotonous behavior: the interval of energies
where phases coexist includes an interval where $d\b(\e)/d\e$ is
positive and the specific heat is negative (see Fig.~5) \cite{us,hu1,gr}.
Hence there is an intrinsic difference between the normal and the 
gravitational first-order phase transitions. 

Even more, normal first-order phase transitions are 
found to replace gravitational first-order phase transitions which occur
in the self-attracting systems that we consider here, if the short-range cutoff
is sufficiently increased.
As was noted in \cite{bm2} for $\a=1$, there is
a critical excluded volume radius $r_c$ above which there is no 
discontinuity
in the entropy vs. energy plot. We observed that this trend is generic
for all $0<\a<3$ and holds for both excluded volume 
and soft potential cutoffs. 
Now the critical cutoff radius $r_c(\a)$ increases with 
increasing $\a$, roughly
varying in value
from below $10^{-3}$ for $\a=1/4$, to above $10^{-1}$ for $\a=5/2$,
respectively.
For a system with a cutoff radius larger than $r_c(\a)$, the entropy vs.
energy plot is
continuous and exhibits all characteristics of a normal first order phase transition 
\cite{us,gr}:
the convex dip and associated with it an
interval of energies, where $d^2s(\e)/d\e^2$ is positive
and the heat capacity is negative (Fig.~5). 
\begin{figure}
\centerline{\epsfxsize=8cm \epsfbox{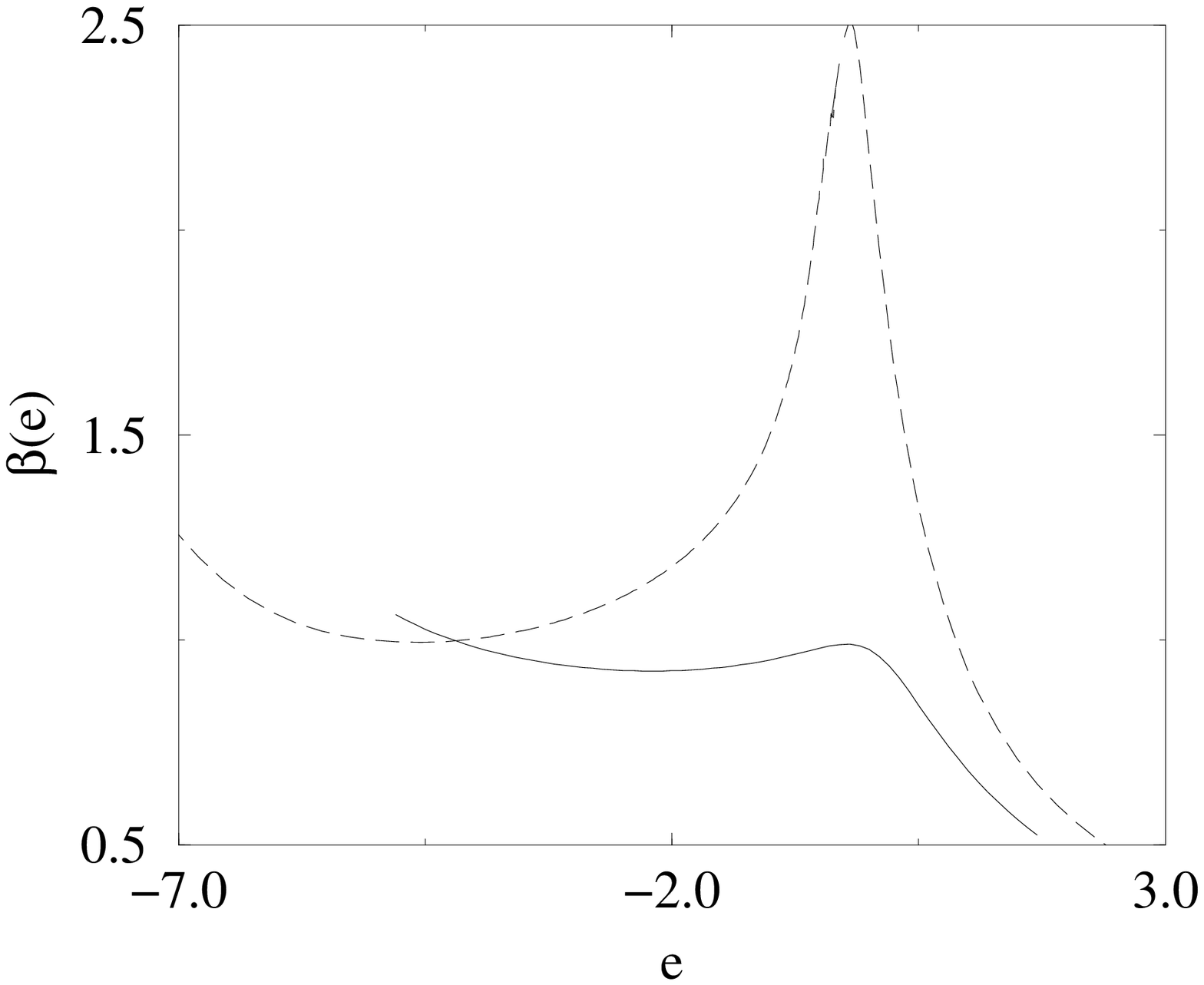}}
\noindent
{\small {\bf Fig.~5.}
Entropy derivative $\b_s(\e)=ds(\e)/d\e$ vs. energy $\e$ plot for
$\a=2$ and central core radius $0.5$ (solid line), and
$\a=1$ and soft potential radius $0.05$ (dashed line).
}
\end{figure}
We leave the more detailed study of the difference between 
the gravitational-like
and the normal first order phase transitions and the
nature of crossover between them
for a future publications.

\section{Acknowledgments}
The authors are grateful to B.~Miller and especially to H.~J.~de~Vega
for helpful discussions.
This work was supported by the Office of Basic Engineering Science
of the U.S. Department of Energy, under Grant No. DE-FG 02-88-ER 13847.

\end{document}